\documentclass[letterpaper,english,reprint,nofootinbib,aps,superscriptaddress,apl]{revtex4-1}

\usepackage{babel,calc,amsmath,amsthm,amssymb,graphicx,,subfigure,xcolor,comment}

\usepackage{array}
\usepackage{txfonts}
\usepackage{amsmath}
\usepackage{graphicx}
\usepackage{dcolumn}
\usepackage{bm}
\usepackage{ulem}
\usepackage{mathdots}
\usepackage{todonotes}
\usepackage[unicode=true]{hyperref}
\usepackage{threeparttable}
\usepackage{times}
\usepackage{CJK}
\hypersetup{
	colorlinks=true,       		
	linkcolor=blue,          	
	citecolor=blue,            
	urlcolor=black,           	
}

\begin{document}

\begin{CJK*}{GB}{gbsn}

\title{Nonlinear manipulation of orbital angular momentum spectra with second- and third- harmonic generation in a quasi-periodically poled crystal}


\author{Yu-Xiang Yang}
\author{Bo-Wen Dong}
\author{Zhi-Cheng Ren}
\author{Hao Li}
\author{Yan-Chao Lou}
\author{Zi-Mo Cheng}
\author{\\ Zhi-Feng Liu}	
\author{Jianping Ding}
\author{Xi-Lin Wang}
\email[]{xilinwang@nju.edu.cn}	
\affiliation{National Laboratory of Solid State Microstructures and School of Physics, Nanjing University, Nanjing 210093, China}
\affiliation{Collaborative Innovation Center of Advanced Microstructures, Nanjing University, Nanjing 210093, China}

\author{Hui-Tian Wang}
\affiliation{National Laboratory of Solid State Microstructures and School of Physics, Nanjing University, Nanjing 210093, China}
\affiliation{Collaborative Innovation Center of Advanced Microstructures, Nanjing University, Nanjing 210093, China}
\affiliation{Collaborative Innovation Center of Extreme Optics, Shanxi University, Taiyuan, China}

\date{\today}


\begin{abstract}
\noindent
Optical orbital angular momentum (OAM), as an important degree of freedom of light, has been attracted extensive attention, due to its intrinsic feature of natural discrete infinite dimension. Manipulation of OAM spectra is crucial for many  impressive applications from classical to quantum realms, in particular, nonlinear manipulation of OAM spectra. Here we realized the nonlinear manipulation of OAM spectra by using the simultaneous second- and third-harmonic generation in a single nonlinear crystal of quasi-periodically poled  potassium titanyl phosphate, for fundamental waves with a variety of OAM spectra, especially for customized OAM spectra of the second and third harmonics. The experimental results confirmed the theoretical predictions. Our approach not only provides a novel way to manipulate OAM spectra at new shorter wavelengths that are hard to be directly generated, but also may find new applications towards multiplexing in classical optics and high-dimensional information processing in quantum optics.
\end{abstract}
	
\maketitle

\end{CJK*}

In 1992, Allen \textit{et al.}~\cite{Allen1992} showed that a Laguerre-Gaussian (LG) beam with a helical phase front of $\exp (j m \varphi)$ can carry an orbital angular momentum (OAM) of $m \hbar$ per photon, where $m$ is the topological charge and $\varphi$ is the azimuth angle in the cylindrical coordinate. Since then, OAM has become one of the most interesting, significant and useful degrees of freedom for photons from fundamental research to practical applications~\cite{Padgett2017}. The topological charge $m$ is valued the arbitrary integer in principle, because such LG beams are the eigen modes of propagation, the spatially structured light with a certain OAM spectral distribution is very interesting and important for a verity of applications, such as spectra self-imaging~\cite{Lin2021}, measuring object parameters~\cite{Xie2017}, optical images rotation and reflection~\cite{Li2018}, multiplexing towards optical communications~\cite{Wang2012, Bozinovic2013} and holography~\cite{Fang2020}, and high-dimensional quantum information processing~\cite{Erhard2020}. The manipulation of OAM spectra becomes a tremendous task for various applications ranging from classical optics to quantum optics and quantum information.  

The OAM spectra can be manipulated based on different principles. Multi-fold rotational symmetry in amplitude distribution for a light beam will tailor its OAM spectrum, such as a three-leaf clover amplitude leads to OAM modes with topological charges being only $3N$ (where $N$ is an integer)~\cite{Guo2022}. Besides the amplitude modulation along the azimuthal direction, inspired by the quasi-mapping relationship between position and OAM spaces, a method using multi-level amplitude ring aperture was proposed and demonstrated to implement the single-step shaping of OAM spectrum~\cite{Pinnell2019}. By modulating the amplitude, another interesting and simple method is proposed to produce a state of light with a controllable OAM spectrum using a binary array of pinholes~\cite{Yang2019}. As the amplitude modulation will limit the manipulation efficiency, to obtain a higher efficiency, the phase-only modulation is a better approach, which could be accomplished by the azimuthal phase modulation~\cite{Zhu2015, Wan2017}.

Besides the above linear methods, nonlinear manipulation is also an important way to control the OAM spectra. Nonlinear up-conversion, such as second-harmonic generation (SHG) and third-harmonic generation (THG), enables to manipulate OAM in the shorter wavelengths without changing topological charges~\cite{Zhou2016, Wu2020, Ren2021, Guo2022, Xu2018}. Based on the OAM conservation law in the nonlinear optical process, the topological charge could be double for SHG~\cite{Dholakia1996, Li2013, Zhou2014, Ni2016} (even for fractional topological charge~\cite{Li2013, Ni2016}) and triple for THG~\cite{Fang2016, Lou2022}. For the inverse process, nonlinear down-conversion is also very significant to manipulate the OAM spectrum, which provides a very useful way to prepare the OAM entangled photon source~\cite{Mair2001}. In the nonlinear interaction, if the involved OAM has a spectrum instead of only single OAM state, we can not only develop a useful way to manipulate OAM spectrum, but also anticipate more novel effects and applications. Therefore, the nonlinear manipulation of OAM spectrum will be an interesting and important topic.

In this Letter, we demonstrate the nonlinear manipulation of OAM spectra by the simultaneous SHG and THG in a quasi-periodically poled potassium titanyl phosphate (QPPKTP). The details of QPPKTP could be found in our previous work~\cite{Lou2022}. The quasi-periodic structure of second-order nonlinear coefficient in QPPKTP provides two pertinent reciprocal vectors supporting two coupled second-order nonlinear processes. One is the SHG of the fundamental wave (FW) to produce the second harmonic (SH) and the other is the sum-frequency generation (SFG) of the FW and SH to generate the third harmonic (TH). These two second-order nonlinear processes happen simultaneously and are coupled with each other, we could implement the nonlinear manipulation of OAM spectra by using SHG and THG in one single nonlinear crystal. 

\begin{figure*}[!ht]
	\centering
	\includegraphics[width=0.9\linewidth]{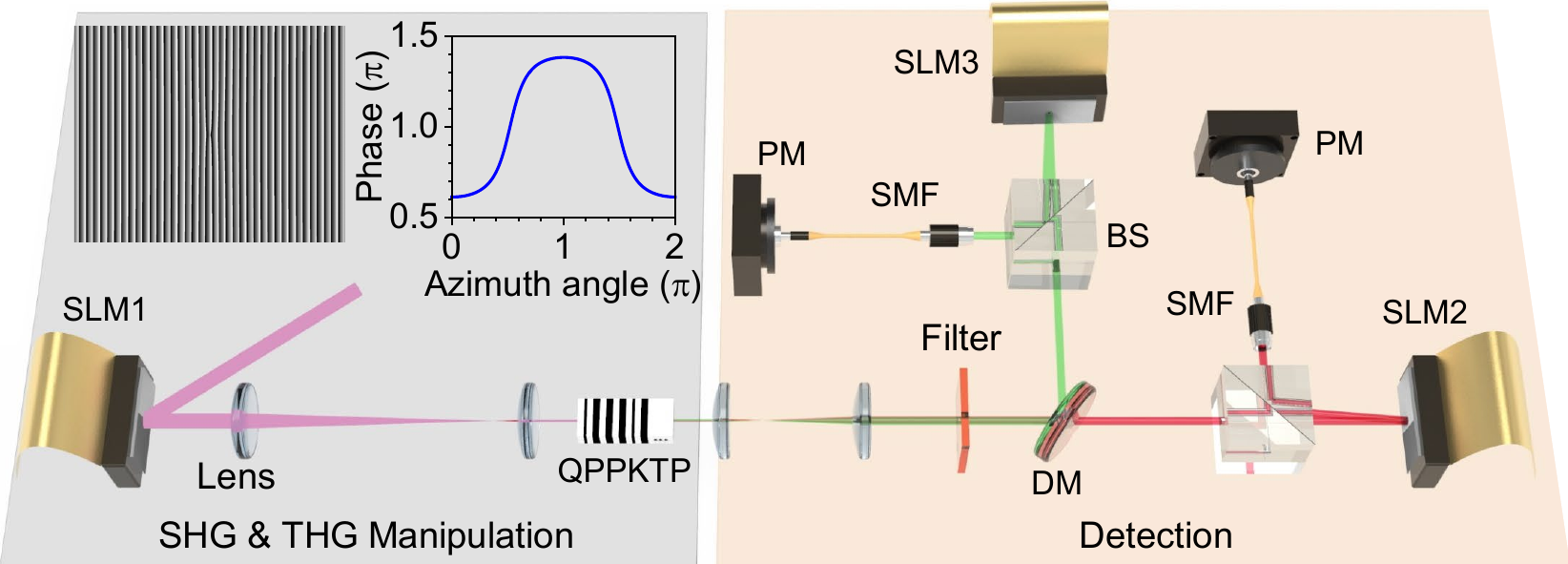}
	\caption{Experimental setup for nonlinear manipulation and detection of OAM spectrum. DM: dichroic mirror, BS: beam splitter, SMF: single mode fiber, PM: power meter.}
	\label{fig:1}
\end{figure*}
Our QPPKTP is designed to be a Type-0 (eee) phase-matching scheme at the FW wavelength of $\lambda_{\omega} \! = \! 1560$ nm. For an input FW of $E_{\omega} \! = \! A_{\omega} \exp(j \psi_{\omega})$ in the QPPKTP, where $A_{\omega}$ is the amplitude and $\psi$ is the spatially-varying phase, under the phase-matching condition, the generated SH should have the form of
\begin{equation}
E_{2\omega} \propto E_{\omega}^{2} = A_{\omega}^2 \exp(j 2 \psi_{\omega}).
\end{equation}
The TH generated from SFG of the FW and SH should have the form as  
\begin{equation}
E_{3\omega} \propto E_{2\omega} E_{\omega} = A_{\omega}^3 \exp(j 3 \psi_{\omega}).
\end{equation}
We can see that the SH and TH have the spatially-varying phases of $\psi_{2 \omega} = 2 \psi_{\omega}$ and $\psi_{3 \omega} = 3 \psi_{\omega}$, respectively, which are of great importance. In this Letter, we devote to only the situation that the spatially-varying phase is azimuthally-varying, i.e.~$\psi_{\omega} \! \rightarrow \! \psi_{\omega} (\varphi)$. When the FW has a helical phase, the double (triple) relationship between the SH (TH) and FW phases clearly understand the topological charge doubling and tripling effects in SHG~\cite{Dholakia1996, Li2013, Zhou2014, Ni2016} and THG~\cite{Fang2016, Lou2022}. Generalizing the helical phase with a single OAM mode to a more complicated phase with an OAM spectrum, the phase doubling and tripling in SHG and THG will provide a novel way to manipulate the OAM spectrum at new wavelengths.      

To demonstrate the above proposal, we construct an experiment setup as shown in Fig.~\ref{fig:1}. The FW is prepared by a femtosecond (fs) pulsed laser with a repetition rate of 80 MHz, a pulse width of 200 fs and a central wavelength of 1560 nm, and its spatially-varying phase is modulated along the azimuthal direction by a spatial light modulator (SLM) as $\psi_{\omega}(\varphi)$. Then the FW is incident into the QPPKTP after its size being reduced to 1/15 by a pair of lenses. The OAM spectrum of the generated SH (TH) is analyzed by the projection measurement with the combination of the SLM2 (SLM3), a single-mode fiber and a power meter. 

\begin{figure}[!ht]
	\centering
	\includegraphics[width=\linewidth]{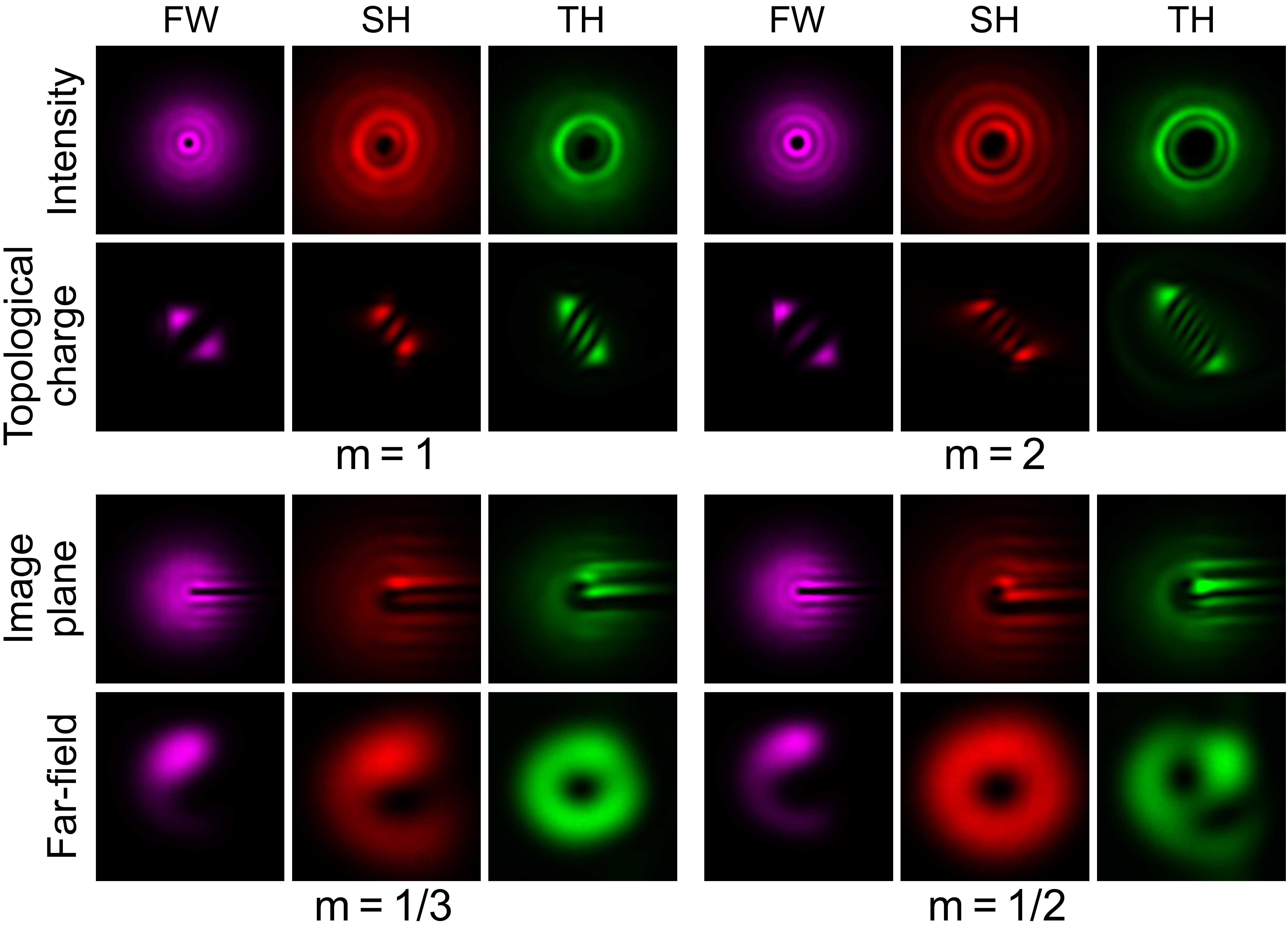}
	\caption{Experimental results of the SH and TH generated by the FW with the integer ($m \! = \! 1$, $m \! = \! 2$) and fractional ($m \! = \! 1/3$, $m \! = \! 1/2$) topological charges.}
	\label{fig:2}
\end{figure}  

\begin{figure}[!ht]
	\centering
	\includegraphics[width=\linewidth]{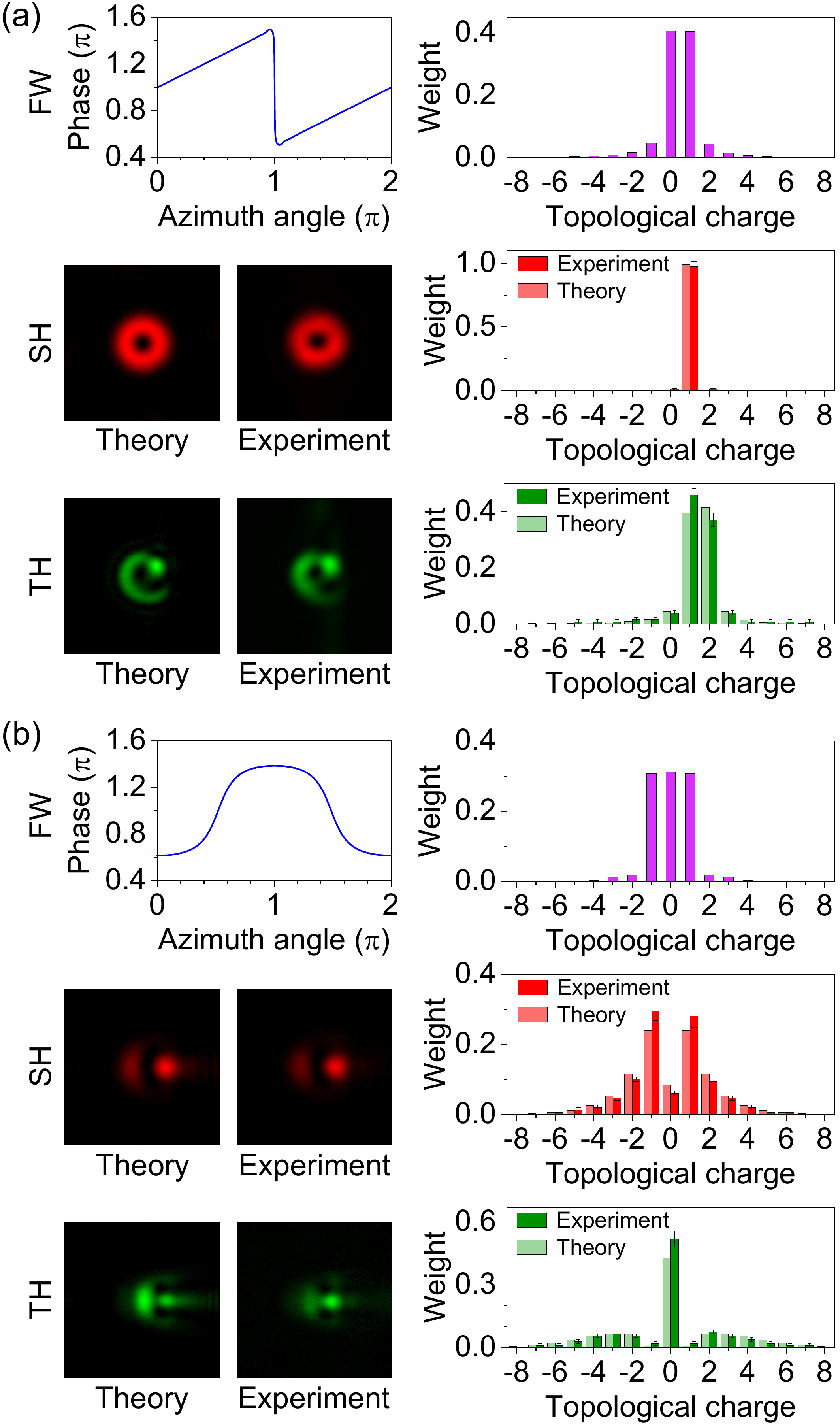}
	\caption{Experimental results for nonlinear manipulation of OAM spectra for the FW with the OAM spectral distribution. (a) $W_\omega^{0} \! : \! W_\omega^{1} \! = \! 1 \! : \! 1$ and (b) $W_\omega^{-1} \! : \! W_\omega^{0} \! : \! W_\omega^{1} \! = \! 1 \! : \! 1: \!1$. The fidelities of the measured OAM spectra with respect to the theoretical predictions are (a) $0.98 \! \pm \! 0.02$ for the SH and $0.98 \! \pm \! 0.02$ for the TH, (b) $0.99 \! \pm \! 0.02$ for the SH and $0.97 \! \pm \! 0.03$ for the TH.}
	\label{fig:3}
\end{figure}  

\begin{figure*}[!ht]
	\centering
	\includegraphics[width=0.99\linewidth]{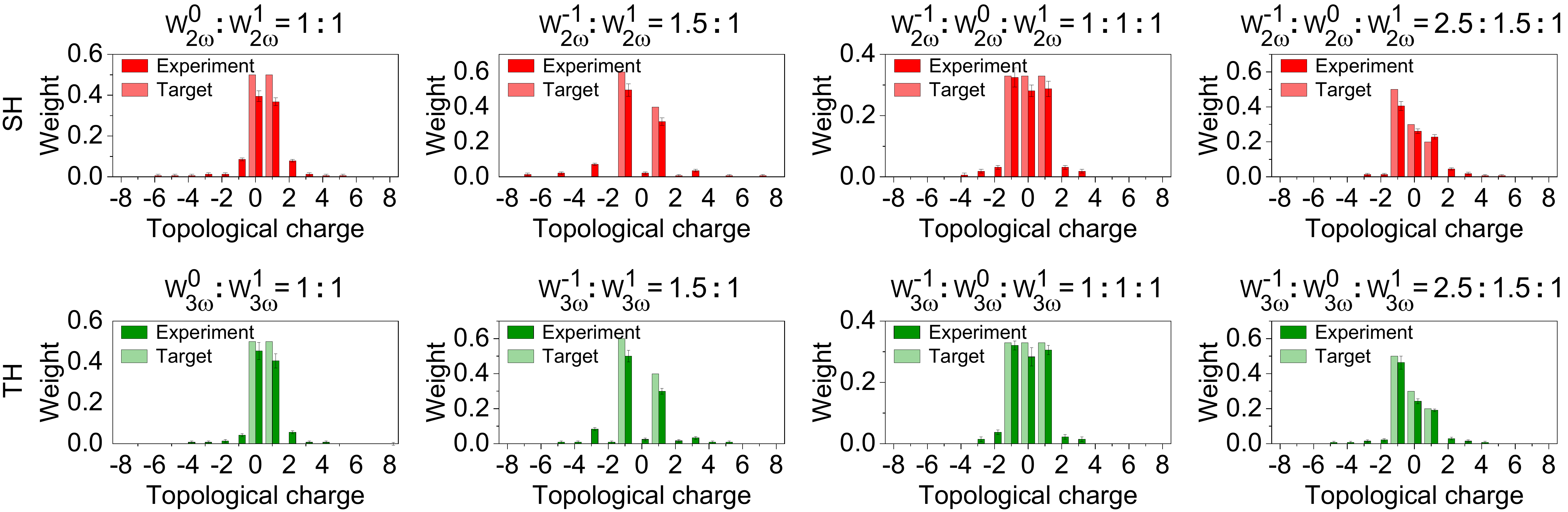}
	\caption{Experimental results for nonlinear manipulation of OAM spectra for the four targets with the predesigned OAM spectra of the SH and TH. The fidelities of the experimentally generated OAM spectra with respect to the targets of the SH and TH are $0.87 \pm 0.02$ and $0.93 \pm 0.03$ for $W_{2 \omega (3 \omega)}^{0} \! : \! W_{2 \omega (3 \omega)}^{1} \! = \! 1 \! : \! 1$; $0.90 \pm 0.02$ and $0.89 \pm 0.02$ for $W_{2 \omega (3 \omega)}^{-1} \! : \! W_{2 \omega (3 \omega)}^{1} \! = \! 1.5 \! : \! 1$; $0.94 \pm 0.02$ and $0.95 \pm 0.02$ for $W_{2 \omega (3 \omega)}^{-1} \! : \! W_{2 \omega (3 \omega)}^{0} \! : \! W_{2 \omega (3 \omega)}^{1} \! = \! 1 \! : \! 1 \! : \! 1$; $0.94 \pm 0.02$ and $0.95 \pm 0.02$ for $W_{2 \omega (3 \omega)}^{-1} \! : \! W_{2 \omega (3 \omega)}^{0} \! : \! W_{2 \omega (3 \omega)}^{1} \! = \! 2.5 \! : \! 1.5 \! : \! 1$.}
	\label{fig:4}
\end{figure*}

Firstly, we prepare the FW with the phase as 
\begin{equation}
\exp (j \psi_{\omega})=\exp (j m \varphi).
\end{equation}
Correspondingly, the generated SH and TH will have the double and triple phases of 
\begin{align}
\exp (j \psi_{2 \omega})=\exp (j 2 m \varphi),\\
\exp (j \psi_{3 \omega})=\exp (j 3 m \varphi).
\end{align}
The experimental results for the FW with two integer topological charges of $m \! = \! 1$ and $m \! = \! 2$ are shown in the first and second rows of Fig.~\ref{fig:2}. The topological charges are verified by counting the petals near the focal plane of a tilt lens and the measured results are in good agreement with the theoretical predictions from Eqs.~(4) and (5).  

A more interesting case is when the FW topological charge is a fraction, such as $ m \! = \! 1/3$ and $1/2$. Correspondingly, the topological charges of the generated SH (TH) should be $2 m \! = \! 2/3$ and $1$ ($3 m \! = \! 1$ and $3/2$), respectively. In theory, these results indicate that the intensity pattern of the generated TH (SH) should form a whole doughnut shape for the FW with $m \! = \! 1/3$ ($1/2$). If we capture the intensity pattern in the image plane, the phase singular line of the FW along the $\varphi=0$ direction will give rise to an extinction line in its intensity pattern, and then results in the similar extinction lines in the intensity patterns of the SH and TH, as shown in the third row of Fig.~\ref{fig:2}. To reveal the difference between them, we also measured the far-field intensity patterns, as shown in the fourth row of Fig.~\ref{fig:2}. As the above theoretical predictions, the intensity pattern  of the SH generated by the FW with $ m \! = \! 1/2$ and that of the TH generated by the FW with $ m \! = \! 1/3$ exhibit clearly a whole doughnut intensity shape.

Secondly, we will explore the nonlinear interaction when the FW has the customized OAM superposition states instead of one single OAM state. So, we need to tailor the OAM spectrum of the FW by a predesigned phase modulation. In principle, for a phase modulated along the azimuthal direction $\psi (\varphi)$, it can be expanded with the eigen OAM states as
\begin{equation}
\exp (j \psi) = {\sum}_{m}c^{m}\exp (j m \varphi),
\end{equation}
where $m$ is an integer. The weight for the eigen OAM state with the topological charge $m$ is defined as 
\begin{equation}
W^{m}= \lvert c^{m} \rvert ^2, \ \ \textrm{with} \ \ {\sum}_{m} W^{m} = 1.
\end{equation}
To achieve the customized OAM spectrum of the FW, defined by the weights $W_\omega^{m}$, here we use the algorithm developed in Refs.~\cite{Zhu2015,Wan2017} to compute the required azimuthally-varying phase $\psi_\omega$ of the azimuthally-invariant amplitude FW. We devote to two kinds of OAM superposition states for the FWs containing two eigen OAM states of $W_\omega^{0} \! : \! W_\omega^{1} \! = \! 1 \! : \! 1$ and three eigen OAM states of $W_\omega^{-1} \! : \! W_\omega^{0} \! : \! W_\omega^{1} \! = \! 1 \! : \! 1 \! : \! 1$. The left column of the first row in Fig.~\ref{fig:3}(a) (Fig.~\ref{fig:3}(b)) shows the computed azimuthally-varying phase $\psi_\omega ^{(1:1)}$ ($\psi_\omega ^{(1:1:1)}$) satisfying $W_\omega^{0} \! : \! W_\omega^{1} \! = \! 1 \! : \! 1$ ($W_\omega^{-1} \! : \! W_\omega^{0} \! : \! W_\omega^{1} \! = \! 1 \! : \! 1 \! : \! 1$). Correspondingly, the right column of the first row in Fig.~\ref{fig:3}(a) (Fig.~\ref{fig:3}(b)) shows the OAM spectrum of the FW, estimated from $\psi_\omega ^{(1:1)}$ ($\psi_\omega ^{(1:1:1)}$). Clearly, the computed $\psi_\omega ^{(1:1)}$ ($\psi_\omega ^{(1:1:1)}$) satisfies the requirements of the two customized OAM superposition states. The corresponding azimuthally-varying phases of the generated SH (TH) are $2 \psi_\omega ^{(1:1)}$ ($3 \psi_\omega ^{(1:1)}$) and $2 \psi_\omega ^{(1:1:1)}$ ($3 \psi_\omega ^{(1:1:1)}$). For the FW with $W_\omega^{0} \! : \! W_\omega^{1} \! = \! 1 \! : \! 1$ ($W_\omega^{-1} \! : \! W_\omega^{0} \! : \! W_\omega^{1} \! = \! 1 \! : \! 1 \! : \! 1$), as shown in the second and third rows of Fig.~\ref{fig:3}(a) (Fig.~\ref{fig:3}(b)), the measured OAM spectra and the intensity patterns of the generated SH and TH are in good agreement with the corresponding theoretically simulated results, respectively. 

To quantitatively characterize the experimental results, we use the fidelities of the experimentally measured weights $W_{E}^{m}$ with respect to the theoretically predicted weights $W_{T}^{m}$ in OAM spectrum as~\cite{He2017}  
\begin{equation}
F = {\sum}_{m} \sqrt{W_{E}^{m} W_{T}^{m}}.
\end{equation}
The calculated fidelities of the experimental results of OAM spectra with respect to the theoretically predicted OAM spectra for the SH (TH) are $0.98 \! \pm \! 0.02$ ($0.98 \! \pm \! 0.02$) for the FW with $W_\omega^{0} \! : \! W_\omega^{1} \! = \! 1 \! : \! 1$ and $0.99 \! \pm \! 0.02$ ($0.97 \! \pm \! 0.03$) for the FW with $W_\omega^{-1} \! : \! W_\omega^{0} \! : \! W_\omega^{1} \! = \! 1 \! : \! 1 \! : \! 1$, which further confirm that the experimental results meet the theoretical expectations.  

Another more important task for nonlinear manipulation of OAM spectrum is how to achieve the customized OAM spectra at the new wavelengths of the SH and TH, for example, to realize a flat OAM spectrum with $W_{2 \omega (3 \omega)}^{0} \! : \! W_{2 \omega (3 \omega)}^{1} \! = \! 1 \! : \! 1$ for the SH (TH). Firstly, we still use the algorithm developed in Refs.~\cite{Zhu2015,Wan2017} to compute the target azimuthally-varying phase $\psi_{2 \omega}$ ($\psi_{3 \omega}$) of the azimuthally-invariant amplitude SH (TH). Based on Eqs.~(1) and (2), we can use the inverse calculating method to give the corresponding phase function of the FW as $\psi_{\omega} = \psi_{2 \omega}/2$ ($\psi_{\omega} = \psi_{3 \omega}/3$) for the SH (TH) satisfying the requirement $W_{2 \omega (3 \omega)}^{0} \! : \! W_{2 \omega (3 \omega)}^{1} \! = \! 1 \! : \! 1$. Thus, we can produce the FW meeting requirement to generate the flat OAM spectrum of with $W_{2 \omega (3 \omega)}^{0} \! : \! W_{2 \omega (3 \omega)}^{1} \! = \! 1 \! : \! 1$ for the SH (TH). As shown in the first (second) row of the first column in Fig.~\ref{fig:4}, the measured OAM spectrum of the SH (TH) has a fidelity of $0.87 \! \pm \! 0.02$ ($0.93 \! \pm \! 0.03$) with respect to the corresponding target. 

The second to fourth columns of Fig.~\ref{fig:4} show other three groups of experimental results towards the goal to produce three different OAM spectra of the SH (TH) with $W_{2 \omega (3 \omega)}^{-1} \! : \! W_{2 \omega (3 \omega)}^{1} \! = \! 1.5 \! : \! 1$, $W_{2 \omega (3 \omega)}^{-1} \! : \! W_{2 \omega (3 \omega)}^{0} \! : \! W_{2 \omega (3 \omega)}^{1} \! = \! 1 \! : \! 1 \! : \!1$ and $W_{2 \omega (3 \omega)}^{-1} \! : \! W_{2 \omega (3 \omega)}^{0} \! : \! W_{2 \omega (3 \omega)}^{1} \! = \! 2.5 \! : \! 1.5 \! : \! 1$. The fidelities of the experimentally generated OAM spectra with respect to the targets are within a range from $0.87 \! \pm \! 0.02$ to $0.95 \! \pm \! 0.02$. However, it should be noted that if the fidelities of the experimentally generated OAM spectra are calculated with respect to the OAM spectra produced by the inversely calculated phase functions, they will range from $0.94 \! \pm \! 0.02$ to $0.99 \! \pm \! 0.02$. The difference between these two types of fidelities originates from the imperfect phase modulations with respect to the ideal designs. Clearly, the inversely calculated phase modulations are the dominant factor affecting the fidelity. The detailed fidelities of the experimentally generated OAM spectra with respect to the targets are shown in Fig.~\ref{fig:4}. In future, a challenge is to develop a more optimal algorithm to inversely calculate the phase function meeting the customized OAM spectrum, which will enable the better nonlinear manipulation of OAM spectrum towards the target.

In conclusion, we have successfully demonstrated the nonlinear manipulation of OAM spectra by using the SHG and THG in a single nonlinear crystal of QPPKTP. Based on the clear relationships of doubling and tripling for the phase functions of SH and TH with that of the FW, our nonlinear approach in QPPKTP provides a novel way to manipulate the OAM spectra at new shorter wavelengths, especially for customized OAM spectra of the SH and TH. Our results may open up new possibility for applications of OAM spectra in classical optics~\cite{Zhu2019} and high-dimensional quantum information processing~\cite{Erhard2020}.


This work was supported by the National Natural Science Foundation of China (Nos. 11922406, 91750202); National Key R\&D Program of China (Nos. 2019YFA0308700, 2020YFA0309500, and 2018YFA0306200); Program for Innovative Talents and Entrepreneurs in Jiangsu; Key R\&D Program of Guangdong Province (No. 2020B0303010001).


\begin{thebibliography}{10}
	
	\bibitem{Allen1992} L. Allen, M. W. Beijersbergen, R. J. C. Spreeuw, and J. P. Woerdman, Phys. Rev. A. \textbf{45}, 8185 (1992).
	
	\bibitem{Padgett2017} M. J. Padgett, Opt. Express \textbf{25}, 11265 (2017).
	
	\bibitem{Lin2021} Z. Z. Lin, J. Q. Hu, Y. J. Chen, S. Y. Yu, and C. S. Br\`{e}s, APL Photonics \textbf{6}, 111302 (2021).
	
	\bibitem{Xie2017} G. D. Xie, H. Q. Song, Z. Zhao, G. Milione, Y. X. Ren, C. Liu, R.Z.  Zhang, C. J. Bao, L. Li, Z. Wang, K. Pang, D. Starodubov, B. Lynn, M. Tur, and A. E. Willner, Opt. Lett. \textbf{42}, 4482 (2017).
	
	\bibitem{Li2018} F. S. Li, T. Z. Xu,W. H. Zhang, X. D. Qiu, X. C. Lu, and L. X. Chen, Appl. Phys. Lett. \textbf{113}, 161109 (2018).
	
	\bibitem{Wang2012} J. Wang, J. Y. Yang, I. M. Fazal, N. Ahmed, Y. Yan, H. Huang, Y. X. Ren, Y. Yue, S. Dolinar, M. Tur, and A. E. Willner, Nat. Photonics \textbf{6}, 488 (2012).
	
	\bibitem{Bozinovic2013} N. Bozinovic, Y. Yue, Y. X. Ren, M. Tur, P. Kristensen, H. Huang, A. E. Willner, and S. Ramachandran, Science \textbf{340}, 1545 (2013).
	
	\bibitem{Fang2020}  X. Y. Fang, H. R. Ren, and M. Gu, Nat.  Photonics \textbf{16}, 102 (2020).
	
	\bibitem{Erhard2020} M. Erhard, M. Krenn, and A. Zeilinger, ``Advances in high-dimensional quantum entanglement,'' Nature Rev. Phys. \textbf{2}, 365 (2020).
	
	\bibitem{Guo2022} H. X. Guo, X. D. Qiu, S, Qiu, L. Hong, F. Lin, Y. Ren, and L. X. Chen, Photonics Res. \textbf{10}, 183 (2022).
	
	\bibitem{Pinnell2019}J. Pinnell, V. Rodriguez-Fajardo, and A. Forbes, Opt. Express \textbf{27}, 28009 (2019).
	
	\bibitem{Yang2019} Y. J. Yang, Q.Zhao, L. L. Liu, Y. D. Liu, C. Rosales-Guzm\'{a}n, and C. W. Qiu, Phys. Rev. Appl. \textbf{12}, 064007 (2019).
	
	\bibitem{Zhu2015} L. Zhu and J. Wang, Opt. Express \textbf{23}, 247340 (2015).
	
	\bibitem{Wan2017} C. H. Wan, J. Chen, and Q. W. Zhan, Opt. Express \textbf{25}, 15108 (2017). 
	
	\bibitem{Zhou2016} Z. Y. Zhou, Y. Li, D. S. Ding, W. Zhang, S. Shi, B. S. Shi, and G. C. Guo, Light: Sci. Appl. \textbf{5}, e16019 (2016).
	
	\bibitem{Wu2020} H. J. Wu, B. Zhao, C. Rosales-Guzm\'{a}n, W. Gao, B. S. Shi, and Z. H. Zhu, Phys. Rev. Appl. \textbf{13}, 064041 (2020).
	
	\bibitem{Ren2021} Z. C. Ren, Y. C. Lou, Z. M. Cheng, L. Fan, J. P. Ding, X. L. Wang, and H. T. Wang, Opt. Lett. \textbf{46}, 2300 (2021).
	
	\bibitem{Xu2018} Z. Xu, Z. Y. Lin, Z. L. Ye, Y. Chen, X. P. Hu, Y. D. Wu, Y. Zhang, P. Chen, W. Hu, Y. Q. Lu, M. Xiao, S. N. Zhu, Opt. Express \textbf{26}, 17563 (2018).
	
	\bibitem{Dholakia1996} K. Dholakia, N. B. Simpson, M. J. Padgett, and L. Allen, Phys. Rev. A \textbf{54}, R3742 (1996).
	
	\bibitem{Li2013} S. M Li, L. J Kong, Z. C. Ren, Y. N. Li, C. H. Tu, and H. T. Wang, Phys. Rev. A. \textbf{88}, 035801 (2013).
	
	\bibitem{Zhou2014} Z. Y. Zhou, Y. Li, D. S. Ding, W. Zhang, S. Shi, B. S. Shi, and G. C. Guo, Opt. Express \textbf{22}, 23673 (2014).
	
	\bibitem{Ni2016} R. Ni, Y. F. Niu, L. Du, X. P. Hu, Y. Zhang, and S. N. Zhu, Appl. Phys. Lett. \textbf{109}, 151103 (2016).
	
	\bibitem{Fang2016} X. Y. Fang, G. Yang, D. Z. Wei, D. Wei, R. Ni, W. Ji, Y. Zhang, X. P. Hu, W. Hu, Y. Q. Lu, S. N. Zhu, and M. Xiao, Opt. Lett. \textbf{41}, 1169 (2016).
	
	\bibitem{Lou2022} Y. C. Lou, Z. M. Cheng, Z. H. Liu, Y. X. Yang, Z. C. Ren, J. P. Ding, X. L. Wang, and H. T. Wang, Optica \textbf{9}, 183 (2022).
	
	\bibitem{Mair2001} A. Mair, A. Vaziri, G. Weihs, and A. Zeilinger, Nature \textbf{412}, 313 (2001).
	
	\bibitem{He2017} Y. He, X. Ding, Z. E. Su, H. L. Huang, J. Qin, C. Wang, S. Unsleber, C. Chen, H. Wang, Y. M. He, X. L. Wang, W. J. Zhang, S. J. Chen, C. Schneider, M. Kamp, L. X. You, Z. Wang, S. H\"{o}fling, C. Y. Lu, and J. W. Pan, Phys. Rev. Lett. \textbf{118}, 190501 (2017).
	
	\bibitem{Zhu2019} L. Zhu and J. Wang, Front. Optoelectron. \textbf{12}, 52 (2019).
	
\end{thebibliography}
\end{document}